\documentstyle[12pt,epsfig]{article}
\newcommand{\be}{\begin{equation}}  
\newcommand{\ee}{\end{equation}}  
\newcommand{\bear}{\begin{eqnarray}}  
\newcommand{\eear}{\end{eqnarray}}  
\newcommand{\ba}{\begin{array}}  
\newcommand{\ea}{\end{array}}

\newskip\humongous \humongous=0pt plus 1000pt minus 1000pt

\newif\ifdtup

\def\oldreffmt#1{\rlap{[#1]} \hbox to 2\parindent{}}

\def\figfmt#1{\rlap{Figure {#1}} \hbox to 1in{}}  
  
\def\ie{\hbox{\it i.e.}{}}	  
\def\eg{\hbox{\it e.g.}{}}

\def\Tr{\mathop{\rm Tr}}

\def\VEV#1{\left\langle #1\right\rangle}

\def\beq{\begin{equation}}  
\def\eeq{\end{equation}}  
\def\bea{\begin{eqnarray}}  
\def\eea{\end{eqnarray}}  
\def\half{\frac{1}{2}}  
  
\def\bq{\begin{quote}}  
\def\eq{\end{quote}}

\def\half{\frac{1}{2}}       

\relax  

\newdimen\tdim  
\tdim=\unitlength  
\def\bar{\overline}

\begin{document}  
  \pagestyle{empty}  
\begin{titlepage}  
\def\thepage {}    
  \title{ \vspace*{1.5cm} \bf    
Fractal Theory Space:\\
Spacetime of Noninteger Dimensionality }   
\author{  
\bf  Christopher T. Hill$^1$ \\[2mm]   
{\small {\it $^1$Fermi National Accelerator Laboratory}}\\
{\small {\it P.O. Box 500, Batavia, Illinois 60510, USA}}
\thanks{e-mail: 
hill@fnal.gov }\\
}
\date{Oct. 1, 2002}
\maketitle
\vspace*{-9.5cm}
\noindent
\begin{flushright}  
FERMILAB-Pub-02/249-T 
\end{flushright}

\vspace*{10.0cm}  
\baselineskip=18pt  
  
\begin{abstract}  
{\normalsize 
We construct  matter
field theories in a ``theory space''
that is fractal, and 
invariant under geometrical renormalization group 
(RG) transformations. 
We treat in detail complex scalars, 
and discuss issues related to fermions,
chirality, and Yang-Mills gauge fields.
In the continuum limit these models describe physics
in a noninteger spatial dimension which appears
above a RG invariant ``compactification scale,'' $M$.
The energy distribution of
KK modes above $M$ is controlled
by an exponent in a scaling relation of
the vacuum energy (Coleman-Weinberg potential),
and corresponds to the dimensionality.
For truncated-$s$-simplex lattices with
coordination  number $s$ the spacetime dimensionality 
is $1+(3+2\ln(s)/\ln(s+2))$. The
computations in theory space involve 
subtleties, owing
to the $1+3$ kinetic terms,
yet the resulting dimensionalites are equivalent
to thermal spin systems. 
Physical implications are discussed.
} 
\end{abstract}  
  
\vfill  
\end{titlepage}  
  
\baselineskip=18pt  
\renewcommand{\arraystretch}{1.5}
\pagestyle{plain}  
\setcounter{page}{1}

\section{Introduction}  

All quests for organizing principles of physics
beyond the Standard Model, since the classic era of
grand unification in the late 1970's, have involved
extra dimensions.  The foremost example is supersymmetry,
\cite{Ramond},
in which one postulates Grassmanian extra dimensions
and graded extensions of the Lorentz group.
Supersymmetry and bosonic extra dimensions
are essential to the use of string theory
with matter fields as a complete description of all forces, 
including quantum gravity.
Motivated by certain viable limits of string theory,
\cite{lykken}, 
the possibility of extra conventional spatial dimensions
at the $\sim$ TeV scale, possibly accessible
to future colliders, has lately
become the focus of a lot of activity.
Latticization, \cite{wang0},  or ``deconstruction,''
\cite{georgi},  
of compactified extra dimensional 
theories provides an effective
gauge invariant Lagrangian in $1+3$ dimensions
truncated on $N$ KK modes of
scalars, fermions and gauge fields
in $D$ dimensions.
This  has provided a
point of departure for abstracting a new class 
of models based upon the notion of ``theory space'' 
as emphasized by Arkani-Hamed, Cohen and Georgi \cite{georgi}. 

Theory space, without some defining principles,
is an empty concept. 
A key idea we emphasize presently is that theory space
can be endowed with certain abstract geometrical symmetries 
that are essentially renormalization group (RG) transformations.
These transformations are distinct from scale transformations
and truly reflect a geometric structure of the theory.
Geometrical symmetries in continuum extra dimensions therefore become
replaced by the renormalization group in theory space. 
We can thus turn it around and use the RG to 
generate new kinds of geometries.  
In the present paper we study a nontrivial example
of the latter possibility.

In particular, we will borrow
from condensed matter physics certain recursively defined,
or {\em fractal},
lattices to construct classes of new theory spaces. 
These lattices are defined by recursively ``decorating'' a
lattice of coordination number, $s$, by replacing each
site with a simplex of $s$ sites, preserving the
coordination number $s$. This process
is iterated an arbitrary number, $k$, times. It results
in a lattice, which for us describes
a fractal theory space matter field theory. In the 
$k\rightarrow \infty $ limit it describes a continuum 
theory whose properties are determined by certain scaling laws
of the (zero temperature) quantum theory,
analogous to the scaling laws of critical systems at finite
critical temperature.

The key observation is that
the Feynman path integral for these systems is 
invariant under a sequence of
renormalization group (RG) transformations 
that map the $k$th lattice into the $k-1$
lattice.   In the  $k\rightarrow \infty $ limit, this RG invariance
imples a certain scaling law for the vacuum action
functional, \eg, the Coleman-Weinberg 
potential.
This scaling law, and consistency with
the RG transformations imposed as a symmetry, 
leads to the determination of a ``critical exponent,'' $\epsilon$.
This exponent is associated with the number distribution
of KK modes with energy:
\be
N(E) \propto \left(\frac{E}{M}\right)^\epsilon
\ee
Here $\epsilon$ is the dimensionality
of the extra dimensions; $M$ is 
a RG invariant mass scale, which is
interpreted as the effective ``compactification
scale.''
The effects of the extra dimension show up only for
energy scales $E\geq M$ as KK-modes appear.
We will obtain irrational (indeed, transcendental) 
values for $\epsilon$ for
the lattices considered presently.
Since theory space, endowed with
such a geometrical symmetry, is effectively 
a theory of compact extra dimensions in the continuum
limit, we have
thus arrived at a prescription for constructing a
spacetime field theory in a flat  
spacetime of noninteger dimensionality.
Our construction is essentially a regularization
procedure for the theory, which is ultimately defined
as a continuum limit. Certain delicacies associated 
with that limit are discussed, including, e.g., 
the ``Casimir'' effects
associated with compactification.

\begin{figure}[t]    
\vspace{7cm}    
\includegraphics{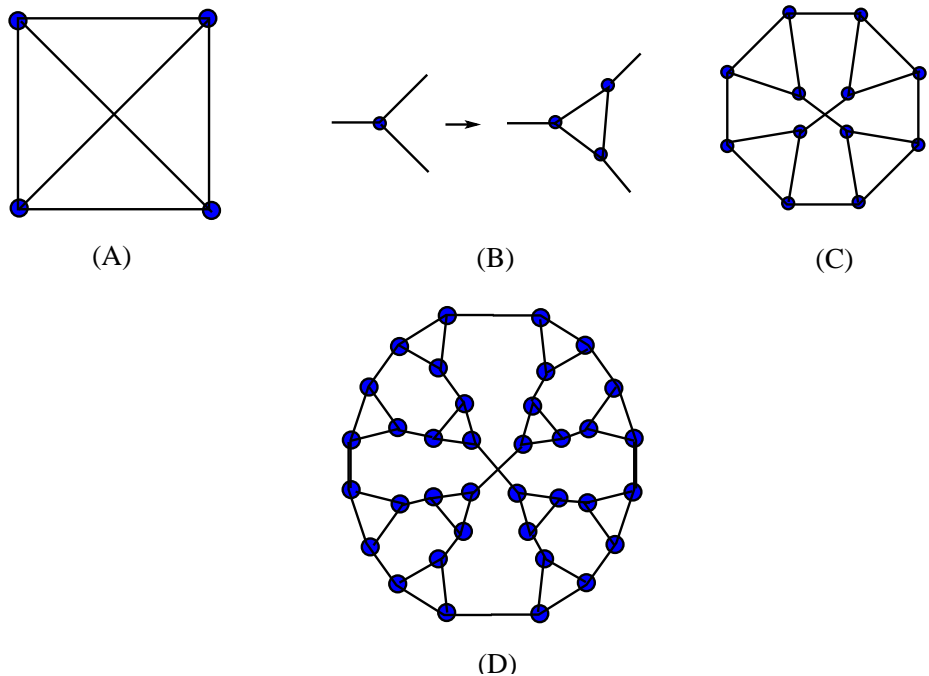}    
\vspace{2cm}    
\caption{\small The truncated $3$-simplex lattice. (A) Kernel 
(complete) lattice  with coordination number $3$; (B) the decoration
which replaces each site under recursion; 
(C) the first order decorated lattice;
(D) the second order decorated lattice. 
A theory space can be constructed by defining each site to
correspond
to a complex scalar $Z|\partial\phi_a|^2 -\mu^2|\phi_a|^2$,
and each link to $-\Lambda^2|\phi_a-\phi_b|^2$.
On the $k$th order truncated $s$-simplex we have the number
of sites $N_k = (s+1)s^k$, and number of links,
$L_k=(s+1)s^{k+1}/2$. 
}   
\label{first}     
\end{figure}    

The kernel lattices we consider are ``complete''
lattices in which every site corresponds to 
a complex scalar field theory of mass $\mu$
in $1+3$ dimensions, and every link is a hopping term,
$\Lambda^2|\phi_a-\phi_b|^2$, 
coupling every scalar field at every site to every other 
site. For example, in Fig.(1) we show a complete
square kernel, equivalent to a tetrahedron, as a zeroth
order kernel lattice with coordination number $s=3$ (this
is called the ``truncated $3$-simplex'').
We then construct the next order lattice by replacing\footnote{
The term ``decorate'' is sometimes used instead of ``replace''
but we will henceforth reserve 
the term ``decorate'' for the RG transformations 
defined below. } 
each vertex with a simplex. This integrates in $s-1$
new fields and $s$ new links per original site. In Fig.(1C)
we have replaced each site of the kernel with  $3$-simplices to
produce the first order lattice. We then iterate 
the replacement to produce the second order lattice
of Fig.(1D). The procedure
can be iterated $k$ times, and we ultimately 
imagine $k\rightarrow \infty$ 
to define a continuum limit. It yields a system
of $N_k =(s+1)s^k$ complex scalar fields coupled through
$L_k = (s+1)s^{k+1}/2$  links.

The renormalization group transformations that define
a symmetry of this system reduce
the $k$th order lattice Lagrangian back to the $(k-1)$th order lattice,
preserving the Feynman path integral.
These  are composed of a sequence of ``polygon-$\star$'' tranformations,
analogous to those first discussed by Onsager for the Ising model
\cite{Onsager},
followed by  ``$4$-chain $\rightarrow$ $2$-chain dedecorations.'' 
These transformations will be adapted to the
$1+3$ field theories that live on sites of the theory
space lattice.  
Such RG manipulations are familiar from the condensed matter
literature, but are tricky in theory space in a
fundamental way: the deconstructed theory possesses 
continuum kinetic terms
for the field theory in the $1+3$ Lagrangian. We must 
include renormalization effects on these kinetic terms,
up to irrelevant
operators that are quartic derivatives, e.g. $(\partial^2\phi)^2/\Lambda^2$.
In particular, $|\partial(\phi_a-\phi_b)|^2$ must
be interpreted as a quartic derivative (we'll see that
discarding this term only affects the high mode number part of the spectrum).  
These irrelevant operators of the derivative expansion
are dropped, and the renormalization of the
relevant $|\partial\phi|^2$ terms is determined.
This renormalization plays a crucial role in the scaling law for the 
Coleman-Weinberg potential.
One obtains the
effective Lagrangian in the $(k-1)$th lattice, with parameters that
are renormalized under the transformations. The consistency of
the RG symmetry, \ie, 
of the invariance of the Feynman path integral,
is realized only for a particular value of 
the dimensionality, $\epsilon$.

The solution to
the problem of extracting $\epsilon$ essentially adapts
the scaling theory of critical
exponents  \cite{Fish2}.
We follow closely the beautiful approach of Dhar
\cite{dhar}, who also discussed many other
lattices, and determined the dimensionality for
finite temperature spin systems. 
The scaling property of the Coleman-Weinberg potential, and
the obtained values of $\epsilon$, depend crucially
upon the recursive construction
of the lattices.
Though the physical systems we
consider are different than  the static spin 
systems considered by Dhar, and we are working in the
zero-temperature quantum theory, we nonetheless recover Dhar's result
for the noninteger dimensionality of the truncated-$s$-simplex lattices of
coordination number $s$:
\be
\epsilon = \frac{2\ln(s)}{\ln(2+s)}.
\ee
Moreover, we find that there are additional RG invariants.
One of these is a mass scale $M$ which plays 
the role of the compactification
scale of the theory, and arises somewhat mysteriously, much like
$\Lambda_{QCD}$ by dimensional transmutation.   Below the scale $M$
the theory is governed only by its zero-modes, and lives in
the $1+3$ dimensions of
the original field theories attached to each site of the lattice.
Above $M$ the ``KK-modes'' begin to appear in the RG invariant
distribution of eq.(1.1), which is the main observable of the theory.
Another RG invariant leads to the classical ``running coupling constant''
relationship in $\epsilon$ dimensions, $g^2 \propto (E/M)^\epsilon$.

When we go
over to theories involving fermions
and Yang-Mills fields 
there are various subtleties. We describe these
mostly qualitatively in Section 4. Despite these
subtleties, it appears that a Standard Model 
generalization can be constructed in $1+(3+\epsilon)$
dimensions.  In the 
Conclusions we will address some options to the 
question of physical interpretation.

\section{Transformations for
Deconstructed Lattice Fields}

We begin by considering transformations
which augment or thin the degrees of freedom
of  $1+3$ theories of many complex scalar fields.
These transformations stem from 
symmetries noticed long ago in the Ising model, 
\cite{Kramers,Onsager,Fish1}.   
In the language of
Ising models a single spin$_1$-link-spin$_2$ combination 
in the Hamiltonian can
always be ``decorated,'' \ie, 
written as a spin$_1$-link-spin${}'$-link-spin$_2$ interaction.
That is, we can ``integrate in'' the new spin${}'$,
or ``decorate'' the original single link. Thus, an $N$-spin
system can be viewed as a $2N$ spin system upon decorating.
The decorations can be arbitrarily complicated, involving many
new spins. Conversely, we can ``integrate out'' or ``dedecorate''
the spins internal to a chain whose endpoint spins are then renormalized
(Fig.(2)).
   
\begin{figure}[t]    
\vspace{4cm}    
\includegraphics{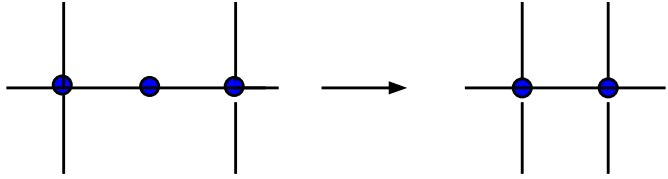}    
\vspace{1.5cm}    
\caption{\small The $3$-chain $\rightarrow$ $2$-chain dedecoration 
transformation integrates out the internal field and renormalizes the 
endpoint fields' kinetic terms and mass terms.}   
\label{first}     
\end{figure}

Decoration is an exact scale transformation for Ising spins, and
continuous spins (e.g., ``spherical models'' are spin systems
which correspond to our models in the absence of kinetic terms). 
Presently we are dealing with a transverse lattice \cite{trans} in which
our ``spins'' are fields that 
have $1+3$ kinetic terms. For us decoration and 
dedecoration
transformations are 
exact scale transformations only in the limit
of very large cut-off $\Lambda$.  This occurs because we perform
decoration transformations truncating on
quartic derivatives, such as $|\partial^2\phi_a|^2/\Lambda^2$.
This is, nonetheless,  a good approximate 
transformation in the $\Lambda\rightarrow \infty$
limit, or for the low lying states in the spectrum. 
These transformations become
symmetries when the theory is classically scale
invariant, \ie, $\mu^2=0$ and $\Lambda\rightarrow \infty$,
and when combined with polygon-$\star$ transformations (below)
on the recursively defined
fractal lattices, they become geometric
symmetries for arbitrary  $\mu^2\neq 0$, \ie, a fixed renormalized
$\mu^2$ can be defined.   
The  
$1+3$ kinetic terms undergo renormalizations
under these transformations, and 
thus distinguish the present construction
from that of a spin model 
(e.g., the continuous complex spherical model).  

We will also require a generalization
of Onsager's ``star-triangle''
or more generally, ``polygon-$\star$'' transformations
that replace a complete polygon of spins, Fig.(3), with
a radiating star configuration, integrating in a new centroid spin.
This transformation can again be done in field theory
to leading order in the derivative expansion,
provided the plaquette
is not oriented (which creates a complication when we
attempt to include fermions and gauge fields). The polygon-$\star$
transformations are, thus, only exact for us in
the $\Lambda\rightarrow \infty$ limit.  There will generally
be hidden symmetries associated with the new centroid field
which may play a role in application to gauge or
chiral--fermion theories.

The key result is that
combining sequences of polygon-$\star$ and dedecoration transformations
allows us to map a recursively defined  
truncated $s$-simplex lattice at $k$-th order into the same lattice
at $k-1$ order with different physical parameters. This leads to
the renormalization group as a geometric symmetry
in the large $k$ limit. The invariance of
the Coleman-Weinberg potential in the large $k$ limit
allows us to determine the dimensionality of the theory.
This limit corresponds to $\Lambda\rightarrow \infty$, and
corrections to the result after truncating the derivative
expansion are vanishing.  The surviving RG invariant parameters
allow us to interpret this as a geometric, as opposed
to a scale, transformation.

\begin{figure}[t]    
\vspace{4cm}    
\includegraphics{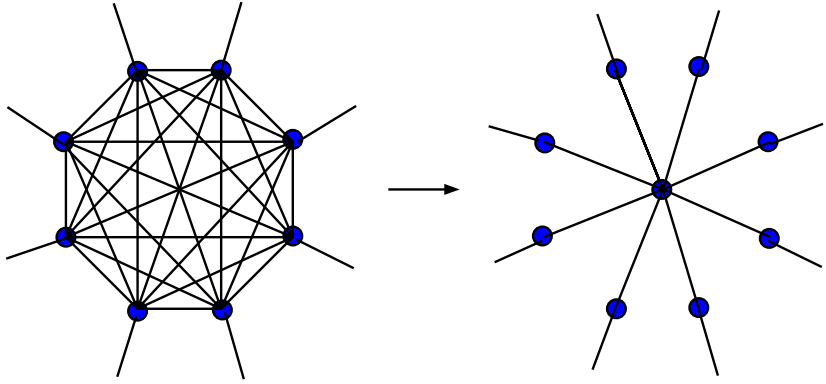}    
\vspace{1.5cm}    
\caption{\small The polygon-$\star$ transformation
for a complete hexagon deletes the intrahexagonal links, 
and integrates in
a new central field with radial links out the  
endpoint fields.}   
\label{third}     
\end{figure}

\subsection{Dedecoration Transformations}
\subsubsection{Example: 3-chains $\rightarrow$ $2$-chains}

We warm up with the simplest example of a ``dedecoration
transformation''  applied to chains of complex scalar fields.
This corresponds to a scale transformation on the theory;
it is a symmetry of the theory only if $\mu\rightarrow 0$.
When $\mu\rightarrow 0$ we will see that the spectra before
and after the dedecoration transformation coincide in
the low energy (low mode number) limit.
 
Consider  an $N$ complex scalar field Lagrangian in $1+3$,
which can be viewed as a deconstructed  $S_1$ compactified extra
dimension  with periodic boundary conditions:
\be
{\cal{L}} = Z_0\sum^{N}_{a=1} |\partial \phi_a|^2 
- \Lambda_0^2 \sum^{N}_{a=1} |\phi_a -\phi_{a+1}|^2 
- \mu_0^2 \sum^{N}_{a=1} |\phi_a|^2 
\label{one1}
\ee
where we take $N$ to be even and assume periodicity, hence
$\phi_{N+a} = \phi_a$. It is convenient to
allow for noncanonical normalization of the kinetic
terms, and we thus display the arbitrary wave-function renormalization
constant $Z_0$.

It is useful to consider
${\cal{L}}$ as a sum over 3-chains:
\be
{\cal{L}} = \sum^{N-1}_{n\;odd} {\cal{L}}_{n,n+2} 
\label{one2}
\ee
Each 3-chain involves three fields. 
The first 3-chain is:
\be
{\cal{L}}_{1,3} = \half Z_0(|\partial \phi_{1}|^2 + 2|\partial \phi_{2}|^2
+|\partial \phi_{3}|^2)
- \Lambda_0^2|\phi_1 -\phi_{2}|^2 - \Lambda_0^2|\phi_2 -\phi_{3}|^2 
- \half \mu_0^2 (|\phi_{1}|^2+2|\phi_{2}|^2+|\phi_{3}|^2)
\label{first}
\ee
The fields $\phi_1$ and $\phi_3$ share
half their kinetic terms and $\mu^2$ terms
with the adjacent chains, thus carry the normalization factors
of $1/2$ within the chain (more generally, the endpoint fields
may have $s-1$ links with other fields and thus carry $1/s$ factors
in the kinetic and mass terms of each chain).
$\phi_2$ can be viewed as a ``decoration'' of the
chain. We can integrate out the internal field $\phi_2$ and obtain
an equivalent renormalized chain.
Integrating out  $\phi_2$:
\be
{\cal{L}}_{1,3} = \half Z_0(|\partial \phi_1|^2 + |\partial \phi_1|^2) - 
(\Lambda_0^2+\half \mu_0^2)(|\phi_1|^2+|\phi_3|^2)
+ \Lambda_0^4 (\phi_1+\phi_{3})^\dagger \frac{1}{Z_0\partial^2 +
2\Lambda_0^2+\mu_0^2}(\phi_1+\phi_{3})
\label{first2}
\ee
Expanding in the derivatives and regrouping
terms gives:
\be
{\cal{L}}_{1,3} = \half Z_1(|\partial \phi_1|^2 + |\partial \phi_3|^2)- 
\Lambda_{1}^2|\phi_1 -\phi_{3}|^2 -\half\mu_{1}^2
(|\phi_1|^2 + |\phi_3|^2)
-\delta_1|\partial (\phi_1-\phi_{3})|^2 + {\cal{O}}(\partial^4/\Lambda^2)
\label{first3}
\ee
where we obtain:
\bea
Z_1 & = & Z_0\frac{8\Lambda_0^4 + 4\Lambda_0^2\mu_0^2 
+\mu_0^4}{4\Lambda_0^4 + 4\Lambda_0^2\mu^2 +\mu_0^4}
\approx 2Z_0
\nonumber
\\
\Lambda_{1}^2 & = & \frac{\Lambda_0^4}{2\Lambda_0^2 +\mu_0^2}  \approx
\half\Lambda_0^2
\nonumber
\\
\mu_{1}^2 & = & \frac{2\mu_0^2 \Lambda_0^2+\mu_0^4}{2\Lambda_0^2 +\mu_0^2} 
\approx \mu_0^2
\nonumber
\\
\delta_1 & = & \frac{Z_0\Lambda_0^4}{(2\Lambda_0^2 +\mu_0^2)^2}\approx
\frac{1}{4}Z_0
\label{renorms}
\eea
We have written the approximate forms of the
renormalizations of the parameters in the large
$\Lambda$ limit.
Note that the $\mu^2$ term is multiplicatively renormalized.
This owes to the fact that it is the true scale-breaking term in the
theory when the lattice is taken very fine, and $\Lambda$ terms 
become  derivatives, \ie, as $\mu\rightarrow 0$ the theory
has a zero-mode. Since it alone breaks the symmetry
of scale-invariance, elevating the zero-mode,
it is therefore multiplicatively renormalized in free field theory.

The $\delta $ term has been written in the indicated form
because, though it superficially appears to be a relevant $d=4$
operator, 
it too is a quartic derivative on the lattice, \ie, ($\partial^2$ in $1+3$)
$\times$(a nearest neighbor hopping term on the lattice). It effects
only the high mass limit of the KK mode spectrum. It is therefore dropped
for consistency with the expansion to order $\partial^4/\Lambda^2$.

The fields develop a new wave-function renormalization
constant $Z_1$. Note that in the $\Lambda>>\mu$ limit, $Z_1\rightarrow 2Z_0$,
{\em twice} the original normalization.  
This renormalization is 
common to all the $\phi_a$ fields
in the other chains. 
Thus, we can write the original theory with
$N$ fields as a sum over
the renormalized 2-chains, containing a total of $N/2$
fields:
\be
{\cal{L}}' = Z_1\sum^{N/2}_{a=1} |\partial \phi_a|^2 
- \Lambda_1^2 \sum^{N/2}_{a=1} |\phi_a -\phi_{a+1}|^2 
- \mu_1^2 \sum^{N/2}_{a=1} |\phi_a|^2 
\label{one4}
\ee
The dropping of the $|\partial(\phi_a-\phi_b)|^2$
terms, which appear to be superficially relevant,
affects only the high energy
spectrum of eq.(\ref{one4}) when $\mu =0$ in
the limit $\Lambda\rightarrow\infty$
and $N\rightarrow\infty$, holding $M=\Lambda/N$ fixed.   
To see this, we diagonalize eq.(\ref{one1}) to
obtain the mass spectrum:
\beq
m^2_0(n) = Z_0^{-1}\Lambda_0^2\sin^2(\pi n/N) + Z_0^{-1}\mu_0^2 \qquad n=(0,1,..,N-1)
\eeq
Diagonalizing eq.(\ref{one4}) yields
the mass spectrum:
\beq
m^2_1(n) =  Z_1\Lambda_1^2\sin^2(2\pi n/N)+ Z_1^{-1}\mu_1^2 \qquad n=(0,1,..,N-1)
\eeq
and comparing with eqs.(\ref{renorms}) we see
that $m^2_1(n)\approx m^2_0(n)$ for $n<<N$, if we neglect
the $\mu^2$ terms. The renormalized mass
term $Z_1^{-1}\mu_1^2 = (1/2)Z_0^{-1}\mu_0^2$, 
is halved
by the dedecoration transformation, and thus we
have performed a scale transformation on the original theory.
The two spectra coincide in the small $n$ limit of
the scale-invariant theory with $\mu^2=0$.

Viewed as a renormalization group, we see that the large
$N$ system flows under repeated application of 
dedecoration transformations to a  block-spin thinned
theory $N' << N$ with $\mu\rightarrow 0$, which is the scale
invariant fixed point.

\subsubsection{Renormalized $4$-chains $\rightarrow$ $2$-chains}

We will require in our applications
presently  the reduction of slightly
more general 4-chains, which
are two endpoint fields and 2 internal decorating
fields.
We must allow for a more general
parameterization
of the chain fields, since this structure
will arise on the lattices of interest {\em after}
a polygon-$\star$ transformation (below). 
Generally, after
performing 
polygon-$\star$ transformations on our
lattice, 
the full Lagrangian will be a sum over
$4$-chains of the form:
\bea
{\cal{L}}^{full} & = & 
\sum_{n}{\cal{L}}_n^{4-chain}
\eea
These $4$-chains will 
live on lattices with a coordination number $s$
and generally have different normalizations for
the two endpoint fields than the two internal fields:
\bea
{\cal{L}}^{4-chain} & = & 
\frac{1}{s} Z_\Phi|\partial \Phi_1|^2+ Z_\phi|\partial \phi_1|^2+Z_\phi|\partial \phi_2|^2
+\frac{1}{s} Z_\Phi|\partial \Phi_2|^2
\nonumber \\ & &
- \frac{1}{s} \mu_0{}^2 (|\Phi_1|^2+s|\phi_1|^2+s|\phi_2|^2+|\Phi_4|^2)
\nonumber \\ & &
- \Lambda'_0{}^2|\Phi_1 -\phi_{1}|^2 - { \Lambda}_0^2|\phi_1 -\phi_{2}|^2 
-  \Lambda'_0{}^2|\Phi_2 -\phi_{2}|^2  
\label{one11}
\eea
The
endpoint $\Phi_i$ fields are  shared with $s-1$ other
neighboring chains, hence the $Z_\Phi/s$ kinetic term
normalization, and the $\mu_0^2/s$ factors.  
Furthermore, note the central
link for the internal fields, $|\phi_1-\phi_2|^2$,
has a different strength ${ \Lambda_0}^2\neq  \Lambda'_0{}^2$
than the extremity links.

We integrate out the internal scalars $\phi_1$ and $\phi_2$.
This requires diagonalizing the $\phi_1$-$\phi_2$ internal
(mass)$^2$ matrix, which has eigenvalues  $\Lambda'_0{}^2$ and
$\Lambda'_0{}^2+2\Lambda_0{}^2$. We then 
regroup the derivative terms
as before, discarding quartic and higher derivatives. We 
thus obtain a renormalized $2$-chain: 
\be
{\cal{L}}^{2-chain} = \frac{\tilde{Z}}{s} (|\partial \Phi_1|^2 + |\partial \Phi_2|^2)- 
\tilde{\Lambda}^2|\Phi_1 -\Phi_{2}|^2 -\frac{\tilde{\mu}^2}{s}(|\Phi_1|^2 + |\Phi_2|^2)
+ {\cal{O}}(\partial^4/M^2)
\label{one12}
\ee
where:
\bea
\tilde{Z} & = & Z_\Phi+ {sZ\Lambda'_0{}^4}\left[
\frac{1}{(\Lambda'_0{}^2 + \mu_0{}^2)^2}
\right]
\nonumber
\\
\tilde{\Lambda}^2 & = & {\half \Lambda'_0{}^4}\left[
\frac{1}{(\Lambda'_0{}^2 + \mu_0{}^2)}-\frac{1}{(\Lambda'_0{}^2 
+2\Lambda_0^2+ \mu_0{}^2)}
\right]
\nonumber
\\
\tilde{\mu}^2 & = & 
\mu_0{}^2\frac{(1+s)\Lambda'_0{}^2+\mu_0{}^2}{\Lambda'_0{}^2 + \mu_0{}^2}
\eea
It is useful to define the ratio $\kappa=\Lambda'_0{}^2/\Lambda_0^2$
and consider the large $\Lambda$ limit of these expressions:
\bea
\tilde{Z} & \rightarrow & {Z_\Phi}+ Zs
\nonumber
\\
\tilde{\Lambda}^2 & = & {\Lambda'_0{}^2}\frac{\kappa}{(\kappa + 2)}
\nonumber
\\
\tilde{\mu}^2 & = & 
\mu_0{}^2(1+s)
\label{param0}
\eea
We will find that $4$-chains arising after polygon-$\star$
transformations on the truncated $s$-simplex lattices 
will have $\kappa=s$.

The full Lagrangian after replacing
the $4$-chains by the $2$-chains 
and summing over all $2$-chains, will take the form:
\bea
{\cal{L}}^{full} & = & 
\tilde{Z}\sum_a|\partial \Phi_a|^2
-\tilde{\mu}{}^2 \sum_a|\Phi_a|^2
- \tilde{\Lambda}{}^2\sum_{a,b}|\Phi_a -\Phi_{b}|^2 
\label{one11}
\eea
Note that when the $2$-chains are summed, the $1/s$
factors disappear in overall kinetic and mass term normalizations.

\subsection{Polygon-$\star$ 
Transformations} 

Let us consider a ``complete'' deconstructed
Lagrangian for a polygon of $s$ sites. 
This is a highly nonlocal structure of Fig.(3)
in which all sites are linked to all other sites with a common
bond strength:
\be
{\cal{L}}^{polygon} = Z_0\sum^{s}_{a=1} |\partial \phi_a|^2 -
\half{\Lambda_0^2}\sum^{s}_{a=1} \sum^{s}_{b=1}|\phi_a -\phi_{b}|^2
-\mu_0^2\sum^{s}_{a=1}|\phi_a|^2
\label{gon1}
\ee
Note that we must
be careful not to double count,
the link $|\phi_a-\phi_b|^2$ in double sums, hence
the factors of $1/2$.
It is interesting
to compute the mass spectrum of the perfect polygon
by itself, going
into the Fourier basis:
\be
\phi_a = \frac{1}{\sqrt{s}}\sum^{s-1}_{k=0} e^{\pi i ka/s} \chi_n;
\qquad \phi_{a + s} = \phi_a
\ee
whence:
\be
{\cal{L}} = Z_0\sum^{s-1}_{k=0} |\partial \chi_k|^2 -
 s\Lambda_0^2 \sum^{s-1}_{k=1} | \chi_k|^2 -\mu_0^2 \sum^{s-1}_{k=0} | \chi_k|^2
\label{gon2}
\ee
Note the sum in the second term begins
at $k=1$, so the mode $k=0$ 
is a zero mode
when $\mu=0$.
Hence, renormalizing $\Lambda^2 =\Lambda_0^2/Z_0$
and $\mu^2 = \mu_0^2/Z_0$, 
the spectrum, consists of $s-1$ degenerate
modes of mass $\sqrt{s\Lambda^2 +\mu^2}$, and the 
single mode of mass  $\mu$ with $k=0$. 

The polygon of $s$ complex scalar fields admits
a transformation which introduces a central 
complex scalar field
$\Phi$ and becomes 
the $s$-star with  $(s+1)$ complex scalar fields.
Let us consider the $\star$ action in the form:
\be
{\cal{L}}^\star = Z_\Phi|\partial \Phi|^2+
Z\sum^{s}_{a=1} |\partial \phi_a|^2 - \Lambda'{}^2 \sum^{s}_{a=1} |\Phi -\phi_{a}|^2
-\mu^2_\Phi|\Phi|^2-\mu'{}^2\sum^{s}_{a=1}|\phi_a|^2
\label{gon3}
\ee
Note that all $|\phi_a-\phi_b|^2$ bonds have been deleted
and we introduce new $|\Phi-\phi_a|^2$ bonds radiating
from the central scalar $\Phi$.

Starting with ${\cal{L}}^\star$, we
integrate out $\Phi$:
\bea
{\cal{L}}^\star  
&=& Z\sum^{s}_{a} |\partial \phi_a|^2  
-(\Lambda'{}^2+\mu'{}^2)\sum^{s}_{a}|\phi_a|^2
\nonumber \\
& & \qquad +\left[
\Lambda'{}^4\sum^{s}_{a,b } \phi_a^\dagger \frac{1}{Z_\Phi\partial^2 +
s\Lambda'{}^2+\mu_\Phi^2}\phi_b + h.c.\right]
\eea
Performing the derivative expansion and reorganizing
terms, we thus recover the polygon form of the Lagrangian:
\bea
{\cal{L}}^\star \rightarrow {\cal{L}}^{polygon}
&=& Z_0\sum^{s}_{a} |\partial \phi_a|^2  
-\half{\Lambda_0^2}\sum^{s}_{a,b }|\phi_a-\phi_b|^2
-\mu_0^2\sum^{s}_{a } |\phi_a|^2
+ {\cal{O}}(\frac{\partial^2}{\Lambda^2})\nonumber \\
\eea
and we have the relations:
\bea
Z_0 & = & Z+Z_\Phi\frac{s\Lambda'{}^4}{(s\Lambda'{}^2+\mu_\Phi^2)^2}
\approx Z + \frac{Z_\Phi}{s} 
\nonumber
\\
\Lambda_{0}^2 & = & 
\frac{\Lambda'{}^4}{(s\Lambda'{}^2 +\mu_\Phi^2)}  
\approx  \frac{\Lambda'{}^2}{s}
\nonumber
\\
\mu_{0}^2 & = & 
\frac{(\mu_\Phi^2 + s\mu'{}^2)\Lambda'{}^2
+\mu_\Phi^2\mu'{}^2}{(s\Lambda'{}^2 +\mu_\Phi^2)}  
\approx \mu'{}^2\frac{(s+1)}{s}
\eea
where the approximate expressions hold in the large $\Lambda$
limit, and are all that we ultimately require to implement the
renormalization group.

Note that we have freedom within the ${\cal{L}}^\star$
Lagrangian to 
vary the ratios $Z_\Phi/Z$ and $\mu_\Phi/\mu'$.
We can for example, choose $Z_\Phi=0$, in which case $\Phi$
is a nonpropagating auxiliary field.   The $\Phi$ field will recover
a kinetic term when subsequent chain transformations are performed.
The $Z_\Phi=0$ case is interesting for Yang-Mills, and corresponds to
``integrating in'' an infinite coupling constant gauge field,
and the infinite coupling will run to a finite value after
subsequent chain transformations. 
Presently we will make the convenient choice
$\mu'{}^2  =  \mu_\Phi^2$, but we do not
specify explicitly $Z_\Phi/Z$.  This
will act as a check on
our result.

We can readily invert the
transformation in
the large $\Lambda $ limit. 
In summary, the polygon Lagrangian:
\be
{\cal{L}}^{polygon} = Z_0\sum^{s}_{a=1} |\partial \phi_a|^2 -
\half {\Lambda_0^2}\sum^{s}_{a=1} \sum^{s}_{b=1}|\phi_a -\phi_{b}|^2
-\mu_0^2\sum^{s}_{a=1}|\phi_a|^2
\ee
can be replaced by the $\star$ Lagrangian:
\be
{\cal{L}}^\star = Z_\Phi|\partial \Phi|^2+
Z\sum^{s}_{a=1} |\partial \phi_a|^2 - \Lambda'{}^2 \sum^{s}_{a=1} |\Phi -\phi_{a}|^2
-\mu'{}^2|\Phi|^2-\mu'{}^2\sum^{s}_{a=1}|\phi_a|^2
\ee
with the choice of parameters $(\Lambda\rightarrow \infty)$:
\bea
 \frac{Z_\Phi}{s} + Z  & = & Z_0
\nonumber
\\
\Lambda'{}^2 & = &  s\Lambda_0^2
\nonumber
\\
\mu'{}^2 & = & \mu_0^2\frac{s}{s+1}
\label{param1}
\eea


\subsection{Combining Polygon-$\star$ and $4$-chain
Transformations to Reduce the Truncated $s$-simplex Lattice}

\begin{figure}[t]    
\vspace{3cm}    
\includegraphics{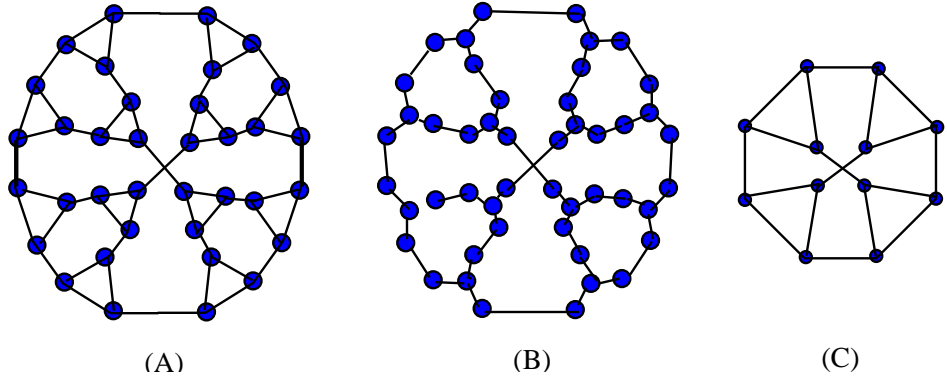}    
\vspace{1.5cm}    
\caption{\small Illustration of $k$$\rightarrow$$k-1$
RG transformation. (A) Second ($k$th) order truncated
$3$-simplex; (B) reduced after polygon(triangle)-$\star$ transformations;
(C) reduced to first ($k-1$) order after $4$-chain
$\rightarrow$ $2$-chain transformations.  }   
\label{third}     
\end{figure}

We are now ready to construct the RG transformation
for the truncated-$s$-simplex lattice by combining
the above transformations. 
Consider any $k$th order $s$-simplex lattice built
up recursively as described in Section I. 
For concreteness
consider Fig.(4A), the second order $3$-simplex.  

The Lagrangian of the $k$th order lattice
takes the form:
\be
{\cal{L}}_k = Z_0\sum^{N_k}_{a=1} |\partial \phi_a|^2 
- \Lambda_0^2 \sum^{L_k}_{links} |\phi_a -\phi_{b}|^2 
- \mu_0^2 \sum^{N_k}_{a=1} |\phi_a|^2 
\label{four1}
\ee
We begin by performing the polygon-$\star$ transformations
on each of the elementary polygons.  All of
the elementary polygons are annihilated by
this procedure, replaced by stars, and the lattice of
Fig.(4A) is carried into that of  Fig.(4B). The centers of
the stars are connected to neighbors through 4-chains,
and the full Lagrangian is now a sum over $4$-chains.
We have the $4$-chain parameters determined
by eq.(\ref{param1}):
\bea
 \frac{Z_\Phi}{s} + Z  & = & Z_0
\nonumber
\\
\Lambda'{}^2 & = &  s\Lambda_0^2
\nonumber
\\
\mu'{}^2 & = & \mu_0^2\frac{s}{s+1}
\eea

Now we reduce the $4$-chains to $2$-chains.  The
lattice is mapped from Fig(4B) into Fig.(4C).
We see that we have now reduced
the original $k$th order lattice to the $k-1$
order with the new Lagrangian:
\be
{\cal{L}}_{k-1} = \tilde{Z}\sum^{N_{k-1}}_{a=1} |\partial \phi_a|^2 
- \tilde{\Lambda}^2 \sum^{L_{k-1}}_{links} |\phi_a -\phi_{b}|^2 
- \tilde{\mu}^2 \sum^{N_{k-1}}_{a=1} |\phi_a|^2 
\label{four1}
\ee
The parameters renormalize as in eq.(\ref{param0})
where $\kappa=\Lambda'{}^2 /\Lambda{}^2 =s$. The
resulting overall renormalization is:
\bea
\tilde{Z} & = &  sZ_0
\nonumber
\\
\tilde{\Lambda}^2 & = & {\Lambda_0^2}\frac{s}{(s + 2)}
\nonumber
\\
\tilde{\mu}^2 & = & 
\mu_0^2s
\nonumber
\\
\tilde{N} \equiv N_{k-1} & = & \frac{N_k}{s}
\label{scaling}
\eea
We have also noted the change in the number
of fields, $N_k$. 
We see that the arbitrariness of choosing ${Z_\Phi}/Z$
(also, $\mu_\Phi/\mu'$, which we fixed to unity) 
in the intermediate step
is a hidden symmetry in the result.

\section{Computation of the Dimensionality}

\subsection{Vacuum Energy Scaling Law}

We have described a theory of free complex
scalars defined on the $k$th
iteration of the kernel lattice. For the
truncated-$s$-simplex lattices
the coordination number is  $s$, the
number of fields in the $k$th iteration is $N_k=(s+1)(s)^k$
and the number of links is $L_{k}=(s+1)s^{k+1}/2$.
The Lagrangian is:
\be
{\cal{L}} = Z\sum^{N_k}_{a=1} |\partial \phi_a|^2 
- \Lambda^2 \sum^{L_{ k}}_{a,b} |\phi_a -\phi_{b}|^2 
- \mu^2 \sum^{N_k}_{a=1} |\phi_a|^2 
\label{dim1}
\ee
where the linking mass term sums over the links.

If we could Fourier transform eq.(\ref{dim1}) we would obtain a
mass spectrum of the form $M_n^2 = \omega^2_n+\mu^2$.
The path integral for our theory then takes the form
in a Euclidean momentum space, up to an overall
multiplicative normalization factor:
\be
e^{-\Gamma} = \int D\phi\; e^{-\int d^4x {\cal{L}}}
= \prod_{p_\mu}\prod_{n=1}^{N_k}(Zp^2 + \omega_n^2 +\mu^2)^{-1}
\ee 
The vacuum energy, or Coleman-Weinberg potential, is up to an overall
additive constant:
\be
\Gamma = Z^{-2}\int \frac{d^4p}{(2\pi)^4} \sum_{n} \ln(p^2 + \omega_n^2 +\mu^2)
\ee
where we have rescaled the $4$ momentum integral
by $Z$.

We want to replace the sum on $n$ by a continuous
momentum integral. In any integer dimensionality,
$n$ is a vector, $\vec{n}=(n_x,n_y,...)$.  We want
to perform the angular integral in
the sum over discrete $\vec{n}$. This leaves 
a sum over the radial 
magnitude of $n=\sqrt{\vec{n}\cdot\vec{n}}$ with
a dimension-dependent measure.
In $\epsilon$ dimensions if we interpret $n$ as
the radial magnitude, then in the continuous
approximation to the sum we can replace:
\be
\sum_{n=0}^{N_k} \rightarrow 
\frac{1}{\epsilon} \int_0^{N_k} dn (n/N_k)^{\epsilon-1}
\ee
In replacing the discrete sum by a continuous
integral we will induce ``Casimir effect''
corrections to the vacuum energy. These
are discussed in the Appendix.

With $n$ a radial coordinate,
the leading behavior at low $n$
of $\omega^2_n$ is $\omega^2_n\sim c(n/N_k)^2\Lambda^2$,
where $c$ is a constant (\eg, $c = \pi^2$ in a
one dimensional periodically compactified situation). 
Let us rescale $n$ to write the integral over
an $4+\epsilon$ dimensional momentum vector, 
\be
p^2 =  p_\mu p^\mu+ c(n/N_k)^2\Lambda^2
\ee 
The Coleman-Weinberg potential becomes:
\be
\Gamma = c(\epsilon)\frac{Z^{-2}N_k }{\Lambda^{\epsilon}}
\int \frac{d^{4+\epsilon}p}{(2\pi)^{4+\epsilon}} \ln(p^2 +\mu^2)
\ee
Here $c(\epsilon)$ is an overall normalizing factor coming from
the $c$ dependence, the ratio of the $d=4$
to $d=4+\epsilon$ solid angles and $(2\pi)^d$
normalizing factors. This factor
is irrelevant for the scaling argument, but given by:
\be
c(\epsilon) = \frac{1}{\epsilon}\left(\frac{4\pi}{c}\right)^{\epsilon/2}
\Gamma(2+\epsilon/2)
\ee 
The integral, apart from the
explicit scaling prefactor,
is finite for nonzero $\epsilon$. It is
thus insensitive to $k$  as $k\rightarrow \infty$ 
to its UV cut-off limit and to
$\Lambda\rightarrow \infty$. We discuss this integral
and the limiting procedure in greater detail
in the Appendix.
Performing the integral:
\be
\Gamma = \frac{Z^{-2} N_k }{\Lambda^{\epsilon}}(\mu)^{4+\epsilon}
V(\epsilon)
\ee
where $V(\epsilon)$ is insensitive to $k$ as $k\rightarrow \infty$.

Thus, suppose we know the value of the parameters $Z$, $\mu$,
$\Lambda$ and $N_k$ for some large value of $k$.
Then, we obtain the Coleman-Weinberg potential for
$k-1$ by the sequence of RG transformations and
we find new parameters:
\be 
\tilde{Z}=h(s)Z,\qquad
\tilde{\mu}=f(s)\mu, \qquad
\tilde{\Lambda}=g(s)\Lambda\qquad 
N_{k-1}=N_k/s.
\ee
Since $V$ is insensitive to $k$ for $k\rightarrow\infty$ (and large $\Lambda$)
we have:
\be
Z^{-2}\frac{N_k }{\Lambda^{\epsilon}}(\mu)^{4+\epsilon} =
\frac{(\tilde{Z})^{-2}N_{k-1} }{(\tilde{\Lambda})^{\epsilon}}
(\tilde{\mu})^{4+\epsilon}
\label{vac}
\ee
or:
\be
1= \frac{(f(s))^{4+\epsilon}}{s (h(s))^2 (g(s))^{\epsilon}} 
\ee
and thus the dimensionality is determined
as:
\be
\epsilon = \frac{(-4\ln(f(s)) + \ln(s)+2\ln(h(s)))}{\ln(f(s)/g(s))} 
\ee

\subsection{Dimensionality and RG Invariants of 
the Trucated-$s$-simplex Lattices}

Let us compute the dimensionality of the truncated $s$-simplex
lattices. From eqs.(\ref{scaling}) we 
 have:
\be
h(s)  =  s, \qquad
f(s)  =  \sqrt{s}, \qquad
g(s)  =  \sqrt{\frac{s}{s+2}}.
\ee
The dimensionality is thus:
\be
\epsilon = \frac{2\ln(s)}{\ln(s+2)}
\ee
We have recovered the result of Dhar
for the dimensionality of spin-systems 
on the truncated $s$-simplex lattices.
In Dhar's analysis of
spins systems, the spins are
static, \ie, have no kinetic terms
in an auxiliary $1+3$ dimensions.  
The wave-function renormalizations
are essential in our present
renormalization group and
to the scaling law for the Coleman-Weinberg potential.
Nonetheless, the lattice dimensionality is the same
as in the static spin system. 

Note that the coordination number must satisfy $s > 2$
for a nontrivial noninteger dimensionality.
For $s=2$ the dimension is always $1$.  For the
truncated $s$-simplex lattices we have $1\leq \epsilon \leq 2$.

The scaling laws amongst the four
quantities of eqs.(\ref{scaling}) imply that
there are $3$ invariants. We have just encountered
one invariant from the vacuum energy scaling law,
and there are thus two others. We list them as follows:
\bea
{V}_r & = & \frac{(\tilde{Z})^{-2}N_{k-1} }{(\tilde{\Lambda})^{\epsilon}}
(\tilde{\mu})^{4+\epsilon}
\nonumber \\
\mu_r^2 & = & \frac{\tilde{\mu^2}}{\tilde{Z}}
\nonumber \\
N_r & = & \tilde{Z}\tilde{N}
\eea
It is useful to define a noninvariant renormalized
cut-off:
\be
\Lambda_r^2 = \frac{\tilde{\Lambda}^2}{\tilde{Z}}
\ee
$\Lambda_r$ can be used as the ``running'' mass scale,
or identified with the energy scale of interest,
$E\sim \Lambda_r$. 

In contrast to the case of the $3$-chain$\rightarrow$$2$-chain
dedecoration acting on a chain of $N$ scalars, the present
RG transformation is {\em not} a scale transformation.
We see that the renormalized mass $\mu_r$ is invariant
under the RG transformation. The present RG transformation is
a statement about the geometric recursive
structure of the theory. This RG invariance of $\mu_r$
emboldens us to consider this as a symmetry of a
novel continuum theory. 

Combining $\mu_r$ with the RG invariant vacuum energy scaling
factor $V_r$ allows us to define yet another
RG invariant mass scale:
\be
M 
= \frac{\tilde{\Lambda}_r}{(\tilde{N})^{1/\epsilon}}
\label{runner}
\ee
The scale $M$ is fixed in the large $\Lambda_r$
and $\tilde{N}$  limit, and has
nothing to do with the physical mass ${\mu}_r$. 
It defines the threshold scale of the
KK-modes, \ie, the effective compactification scale of the theory.
Comparing with the
spherically symmetric measure in the integral
of eq.(3.35) we see
that the number of KK modes with energy $E$ is given by
\be
n(E) = \left(\frac{E}{M}\right)^\epsilon
\ee
$M$ is therefore the ``compactification'' scale of the
noninteger extra dimension.  The scale $M$ persists
in the $\mu\rightarrow 0$ limit. It is somewhat mysterious,
in that we are taking a classical theory to a continuum limit,
yet a nontrivial RG invariant mass scale survives.
It is a consequence of the fact that the dimensionality
is not trivial, \ie, $\epsilon \neq 0$ and a fundamental scale must
survive since the trace of the stress tensor is presumably nonzero
in this classical theory. In this sense, $M$ is analogous to
$\Lambda_{QCD}.$\footnote{$\Lambda_{QCD}$ arises
because the quantum $\beta/g$-function (the Gell-Mann--Low
$\psi$ function) is nonzero and acts
as $\epsilon$, an ``anomalous dimension'' for the coupling constant, 
and hence the trace of the stress tensor is nonzero.} 
For us $\epsilon$
is classical, while in QCD $\epsilon\sim \beta/g$ is an
anomalous dimension.

The physical significance of the invariance of
$\tilde{Z}\tilde{N}$ pertains to interacting theories,
such as Yang-Mills gauge theories.
We can identify:
\be
\tilde{Z} \propto \frac{1}{g^2}
\label{cc}
\ee 
a common dimensionless coupling
constant of the deconstructed theory defined
at the scale of the cut-off. Then the
invariant $M$ tells us how the coupling constant scales
with choice of cut-off. To see the running coupling constant
scaling law
combine
eq.(\ref{runner}) and eq.(\ref{cc}) to obtain:
\be
{g^2(E)}\propto \left(\frac{E}{M}\right)^\epsilon
\label{runcc}
\ee
Thus, we recover normal power-law running
of $g^2(\Lambda_r)$ when $\epsilon$ takes on integer
values. The formula exhibits the generalization
for noninteger dimension.

We have made a continuous approximation to
the integrals, but we can always ``rediscretize'' the sum
in an RG invariant way.  The original
choice of the definition of the energy, $\omega_n
=\sqrt{c}n\Lambda/N_k$ is not RG invariant. We see that the
overall coefficient of the Coleman-Weinberg potential
in eq.(3.36) can be written as:
\be
\frac{Z^{-2}N_k}{\Lambda^\epsilon}(\mu)^{4+\epsilon} =
(\mu_r)^{4+\epsilon}M^{-\epsilon}
\ee
Hence, rediscretizing, we replace
\be
M^{-\epsilon}\int d^{4+\epsilon}p \; I(p)
\sim \int d^{4}p \sum_n (n)^{\epsilon-1} I(p_\mu, \omega_n)
\ee
and the energy of the $n$th mode is
$\omega_n= \sqrt{c}nM$.

The difference between the rediscretized sum, taken to 
infinity, and the continuous integral, is a Casimir effect.
In the Appendix we show that the Casimir effect
can be expressed as a  finite integral.  The finite integral
is vanishing when we assume that the theory can be
expressed in a $4+\epsilon$ dimensional Lorentz invariant way.
It is not necessary to make this strong assumption; the effects
of the noninteger extra dimension may show up in an RG
invariant way only at the threshold
scale $M$, but have bad non-RG invariance at higher energies.
This depends upon the physical interpretation of the
theory. The finiteness of the Casimir integrals suggest
to us that a true RG invariant theory exists, and our
construction for any finite order of $k$ is just a
regulator with RG-symmetry breaking terms.

\section{Considerations of Gauge Fields and Fermions}

Naturally, we are interested in realistic models
built along the lines suggested here. Thus we will require
Yang-Mills, and fermions, including chirality.
The present discussion will be qualitative, as we note some 
new issues that arise in attempting this extension.

When we go
over to theories involving fermions
and Yang-Mills fields 
there are additional subtleties.  These subtleties revolve
around the polygon-$\star$ transformation.
For example, Wilson fermions in a polygon cannot be mapped to the
$\star$ configuration. Similarly, the PNGB's of Yang-Mills theories
that are periodically compactified must be lifted by plaquettes
in order to perform the polygon-$\star$ transformation. The
point is that the polygon is orientable, while the $\star$
is not, so orientational elements of the action 
will not be carried through by the transformation.
In the Yang-Mills case, an arbitrary  magnetic flux threading a
plaquette, $\int B\cdot dA \sim \oint A\cdot dx$ cannot
be represented by the $\star$ form of the action, and this
requires that a certain PNGB be infinitely heavy.
The $\star$ configuration, however, will be seen to be the
key to creating chiral fermions. Chiral fermions in deconstruction
are lattice defects. In the present case they must be incorporated as the
centers of $\star$ configurations that are invariant
under the RG transformations used to reduce the lattice. In a
sense then, chiral fermions are rarified  defects, or 
invariant centers
in the fractal lattice, similar to doping atoms
in a material, or to the centers of snowflakes. 

Yang-Mills gauge fields are introduced in
a deconstructed theory by having
gauge groups, $G_a$, living on sites 
and linking-Higgs-fields defined on links.
We also include plaquette terms which 
show up as mass terms of PNGB's in the $1+3$ dimensions.
Hence, let us choose $G_a = SU(N)$ with a common
coupling constant $g$, and 
the link field $\Phi_{ab}$ is then an $(\bar{N},N)$
chiral field with a VEV, $\VEV{\Phi} = v I_N$.
The Lagrangian is then:
\be
{\cal{L}}_{YM} = -\sum_{n=1}^{N_k} \frac{1}{4g^2} G^a_{\mu\nu}G^a{}^{\mu\nu}
+ \sum_{(ab)}^{L_k} \Tr [D_\mu, \Phi_{ab}]^\dagger[D^\mu ,\Phi_{ab}]
+\sum_{plaq\;n}^{P_k}\lambda_n \Tr[\prod_{plaq\;n}\Phi_{ab}]
\ee
The {\em irreducible }
plaquettes are those which do not encircle a subplaquette
(\ie, can be contracted).  The irreducible plaquettes
are  $P_k=(s+1)^{k+1}$.

\begin{figure}[t]    
\vspace{3cm}    
\includegraphics{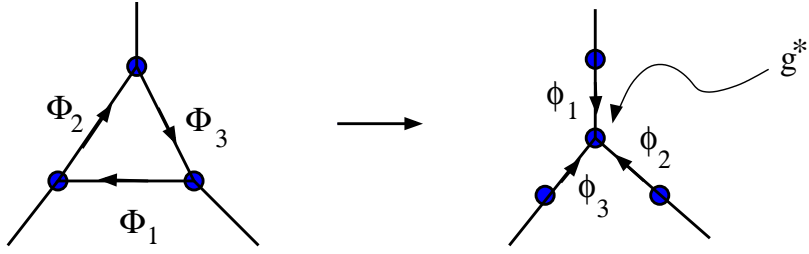}    
\vspace{1.5cm}    
\caption{\small The triangle-$\star$ transformation for an
irreducible plaquette maps $\Phi_i\rightarrow\phi_j$, but
imposes a constraint, $\Phi_3\Phi_2\Phi_1=1$.}   
\label{third}     
\end{figure}

${\cal{L}}_{YM} $ has been supplemented with a plaquette
action, where each plaquette has a coupling
constant $\lambda_k$. Let us first consider $\lambda_k=0$.
Then the theory will contain a spectrum of $1$ vector
zero mode, $N_k-1$ massive
gauge fields (KK-modes), and in tree approximation
$L_k-N_k+1$ massless PNGB's.  The PNGB's will generally be
lifted in perturbation theory to masses of order $\alpha M^2$,
but they can also be elevated by turning on the $\lambda_k$.
Indeed, we see that $P_k>>L_k$, so including all 
irreducible plaquettes
with large $\lambda_n$ we can lift all PNGB's, except
for a single zero-mode.

Lifting the PNGB's is necessary for the implementation
of the $\star$-chain RG.  In Fig.(5) we see a mapping of
the irreducible triangle with link fields $\Phi_i$ into a
star configuration with new link fields $\phi_j$.  The net 
gauge phase rotations in going from one site to another
must be faithfully represented under this redefinition, thus:
\be
\Phi_1 = \phi_3^\dagger\phi_2 \qquad \Phi_2 = \phi_1^\dagger\phi_3
\qquad \Phi_3 = \phi_2^\dagger\phi_1
\ee
and we thus see that the $\Phi_i$ are constrained:
\be
1= \Phi_3\Phi_2\Phi_1
\ee
This is the orientability problem mentioned above. 
It requires the quantization of the Wilson loop
around the triangle plaquette $g\oint A_A dx^A = 2\pi n$
(more properly, $\Phi_3\Phi_2\Phi_1$ must lie in the center
of the group).  In the deconstruction language, it
imposes a constraint on the PNGB's.  We can
treat this constaint by introducing terms
$\sim \lambda_{123} \Tr(\Phi_3\Phi_2\Phi_1)$ for all
elementary plaquettes, and we treat  $\lambda_n$
as a Lagrange multiplier, then
perform the polygon-$\star$ transformation. This will lift
the PNGB's from the spectrum. This is the expected decoupling
that of high mass PNGB's must occur when 
the short-distance degrees of freedom are thinned.
Thus, we expect that polygon-$\star$ transformations should make sense
in the theory with plaquettes.  

An intriguing point is that the $\star$ Lagrangian 
involves ``integrating
in'' additional Yang-Mills gauge groups at the centers
of stars
with coupling constants $g^\star$. As we saw in the
star transformation, there is a freedom to choose
the wave-function renormalization constant, $Z_\Phi$, arbitarily
relative to its neighbors.  This translates into the freedom of choosing the
coupling constant $g^*$ for the new central gauge group arbitrarily.
In particular, we can choose $g^*=\infty$, which completely
suppresses the continuum kinetic term
of the new gauge field at this scale.
The subsequent chain transformations will induce
a finite coupling and gauge invariant kinetic term
for this gauge field as we perform the $4$-chain$\rightarrow$$2$-chain 
transformations. The renormalized couplings after the 
combined transformations for all gauge fields will have a common
value and will run according to the scaling
laws described in the previous section. 

Barring topological obstructions, 
we thus expect that the reduction
for Yang-Mills gauge fields goes through
in the Gaussian approximation. We obtain 
the same dimensionality as for complex scalars.
Obviously the question of the effects
of interactions is of great importance.
We expect that there are $1+3$ continuum renormalization group 
effects that accompany the lattice reduction, which corresponds to
a change of scale (\eg, of $\Lambda_r$). The main issue,
however, comes from the power-law running in eq.(\ref{runcc}).
The Yang-Mills coupling constant as described
classically, will reach evolving upward with scale, 
a unitarity bound, $g^2\sim (4\pi)^2$ at an
effective scale $\Lambda_r^\star$ fairly quickly
(it would be interesting to construct models in which
$\epsilon <<1$ where the power law running is suppressed,
and appears approximately logarithmic). This is the scale
of unitarity breakdown for longitudinal KK-mode
scattering \cite{wang0}. This would
imply a phase transition in the theory,
possibly the string transition. 
Another logical possibility is that $g^2$ runs large,
but then is ``reset'' to a small value by a dynamical
transition in the theory, then runs large again, etc.,
leading to a
{\em limit cycle}.
With a limit cycle it may be
possible to take $\Lambda_r\rightarrow \infty$ in the interacting theory
as well, without a transition to the string phase.
Perhaps the most interesting possibility is 
that the theory has  a UV fixed point
\cite{Keith}, which may arise between a competition
of the classical running and the one-loop correction.

Fermions pose additional challenges. Fermions
live on sites and will have kinetic ``hopping terms''
on the links.  We can always view the lattice as
a fermion mass matrix, take all fermions to be
vectorlike, and choose the hopping terms to be mass terms.
This would readily admit polygon-$\star$ transformations
and RG reduction of the lattice as we have derived.
This would seem to us to be a relatively
uninteresting case.

The hopping terms should be built out of
$\gamma$
matrices.
We expect that we require the use of all $\gamma$ matrices through
$\gamma_{4+[s/2])}$ in construction of the action. 
Hence, for $s=2$, $\gamma_5$ suffices, such as in $S_1$, while $s=3$ requires
$\gamma_6$, etc. Consider the polygon of Fig.(5) for $s=3$
the fermionic hopping terms around the perimeter
of the polygon.  Using 
$\gamma_5$, the hopping terms
are of the form
$\sum_n v\bar{\psi}_n\gamma_5\Phi_n\psi_{n+1}$. Generally
this form leads to the fermion doubling problem
\cite{adam}, but admits polygon-$\star$ transformations. 
It is most sensible
to consider Wilson fermions \cite{adam} 
(the Wilson fermion structure will always occur
with SUSY).  With
Wilson fermions the hopping terms are written
as 
$\sum_n(v\bar{\psi}_{nL}\Phi_n\psi_{(n+1)R} 
- v\bar{\psi}_{nR}\psi_{nR})$. The Wilson
fermion hopping terms have a definite orientational 
sense, $L(n)\rightarrow R(n+1)$,
around the polygon.  These cannot be reduced by polygon-$\star$
transformations, and it is not clear to us that sensible reducible
fermionic actions exist which
are compatible
with the polygon-$\star$ transformation.
It should be borne in mind,
however, that the polygon-$\star$ transformation
is ultimately  a
convenience in computing the dimensionality of
the lattice. More exotic fermionic reductions that 
do not require the polygon-$\star$ transformation may exist.

\begin{figure}[t]    
\vspace{4cm}    
\includegraphics{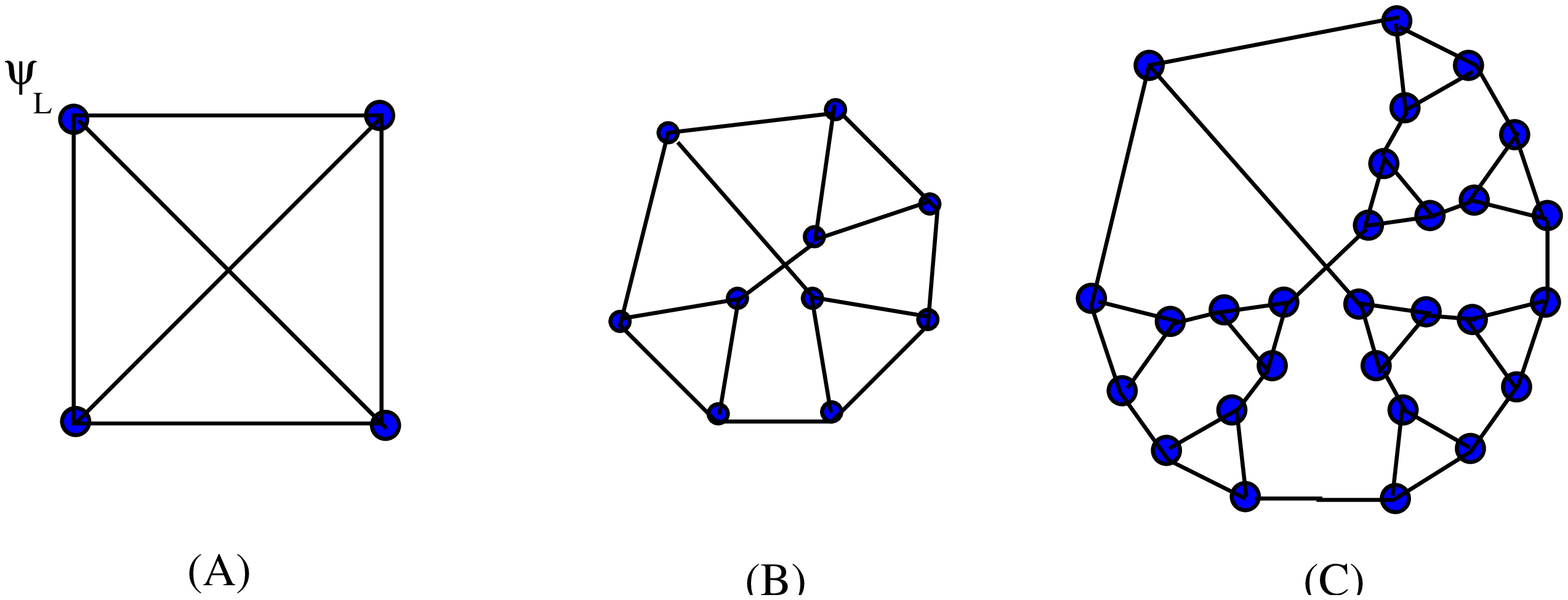}    
\vspace{2cm}    
\caption{\small The truncated $3$-simplex lattice
constructed with an invariant star dislocation
for chirality. (A) Kernel 
(complete) lattice  with coordination number $3$ and
a single chiral fermion attached to the upper lh
site; (B) the first order decoration
replaces each site under recursion except
the invariant star; 
(C) the second order decorated lattice.
A theory space can be constructed by defining each site to
correspond
to a vectorlike
Standard Model, except the invariant star which is
the usual chiral Standard Model. Hopping terms involve
gauge links and Dirac fermion mass terms, $\bar{\psi}_a\psi_b+h.c.$.
The chiral fermions in the invariant star have chiral
hops $\bar{\psi}_{L}\psi_{bR}+h.c.$ to the vectorlike neighbors.
The zero-mode structure of the theory is the chiral
Standard Model. 
}   
\label{first}     
\end{figure}    

If we use an action with vectorlike
fermions and fermionic Dirac mass matrix hopping terms, 
$\bar{\Psi}_a\Psi_b + h.c.$, we can still introduce
chirality. We must construct ``invariant stars,'' \ie, 
dislocations in the lattice that are not reduced
by the RG transformations as  in Fig.(6).
At the center of the invariant star 
configuration we introduce a single chiral fermion, $\Psi_L$. 
The fermion has radial hopping
terms to the perimeter fermions of the form
$\sum_nv\bar{\Psi}_L\phi_n\psi_{nR}$. By ``doping'' a 
mass-matrix lattice
with the appropriate number of 
chiral dislocations one can construct
a fractal imbedding of the Standard Model.

\section{Physical Interpretations and Conclusions}

How do we interpret these new theories physically?
Fractional extra dimensions are not 
obviously compactified extra dimensions,
since no global boundary condition is introduced which corresponds to 
a global compactification.  
Rather, we introduced initially a regulator, $\Lambda$,  
which is our (inverse) 
lattice scale.
We ultimately imagine 
the limit $\Lambda\rightarrow \infty$, but 
how this limit is taken
is dependent upon the physical interpretation of
the theory.   
The analogue of a compactified
theory emerges with the determination
of a compactification scale,
the RG invariant $M$, and is a physical scale held fixed in
the $\Lambda\rightarrow \infty$ limit.  
$M$ would still be present, however,
with a different definition of the theory 
in which we maintain  a finite $\Lambda$,
and we may have a hierarchy $M<<\Lambda$,
in analogy to the usual 
compactified extra dimensional theories.

There are thus two
physical interpretations for these constructions.
The first is an ``outer'' modification of spacetime. Here
we have in mind finite $\Lambda$
but a dimensional transition at the scale $M<<\Lambda$
in which we view the continuous $1+3$ 
dimensions as  {\em a brane in a higher dimension
with a surface structure} 
with characteristic scale length $1/M$. 
 This brane surface is viewed as dynamical, analogous
 to surfaces
in condensed matter physics, and may arise from an interface
with an exterior region involving new physics. The fractal
theory space is an effective description of such a system
on scales not far above $M$. The
fractality, in analogy with surface layers on material media,
may arise because  the interface with the
extra outside dimensions involves a region
of rapid change in physical parameters.
In this picture Lorentz invariance at short distances strictly only
applies to the $1+3$ dimensions, but with $\Lambda$
large the relevant low energy physics of the dimensional
transition scale $M$ is approximately Lorentz invariant
in $1+(3+\epsilon)$ dimensions.
In this case, the scale $\Lambda$ may represent a further
higher energy transition to string theory. Terms of order
$\Lambda^4$ and $\Lambda^2\mu^2$ reflecting the
finite cut-off will be present in the
vacuum energy, and are non-RG invariant.
The low energy
physics, however,  is a fixed point under 
these  renormalizations 
that drastically change the UV part of the theory.

The alternative, and perhaps more intriguing view, is an
``inner'' modification of physics, in that
the scale $M$ represents a true dimensional transition to
$1+(3+\epsilon)$ dimensions, with enhanced Lorentz invariance
on scales above $M$.  In this case we take $\Lambda\rightarrow \infty$.
Then at all shorter distances the noninteger dimensionality is
preserved. This is a remarkable possibility in that a quantum field theory
defined in $1+(3+\epsilon)$ dimensions with
irrational $\epsilon$ is finite to all orders
in perturbation theory. The cut-off scale
$\Lambda$ can be taken to infinity with impunity, 
holding $M$ fixed as the defining
dimensional transition scale. 

If we were naive, we would speculate that we
have given a prescription for the construction of
{\em finite} quantum field theories of matter.
Thus, all infinites in $1+3$ dimensions
of the Standard Model would be associated with the
cut-off scale $M$, which  is  the threshold for new physics
associated with the noninteger correction to the dimension
of space-time. Above the scale $M$ we would begin
to include KK-modes in accord with the energy
distribution $(E/M)^\epsilon$, and we would treat the
field theory with `t Hooft--Veltman dimensional
regularization as the exact calculational tool
for $4+\epsilon$ dimensions. Thus,
one way to treat the Standard Model as a quasi-noninteger
dimensional theory would be to replace all loop integrals
in $4$-dimensions by $4+\epsilon$ dimensions above
a fixed matching scale $M$:
\be
\makebox{Loops} \rightarrow \int_0^M\frac{d^4k}{(2\pi)^4} 
+
M^{-\epsilon}\int_M^\infty\frac{d^{4+\epsilon}k}{(2\pi)^4} 
\ee
Modulo ambiguities in treating $\gamma^5$,
This theory is apparently finite above $M$ to 
all orders in perturbation theory. 
We would infer some immediate physical
consequences, \eg, 
that the Higgs boson will receive 
radiative corrections to its mass from top quark loops:
\be
m_H^2 \sim -\frac{3g_t^2}{8\pi^2} M^2 
\ee
(or heavier fermions in an extension of the model).
Hence, we infer that $M^2$ is of order
$\sim$ TeV (a ``Little Higgs'' model can
raise the scale to $\sim$ 10 TeV through
custodial chiral symmetries).  

However, the power-law running of
the coupling constant, $g^2\sim (E/M)^\epsilon$
with $\epsilon>0$, 
implies that either (1) the theory
has a UV fixed point or (2) has a limit cycle
or (3) undergoes a phase transition at
a strong coupling scale $g^2 \sim (4\pi)^2$. 
 In any case
we must account for gravity, and imbedding into string
theory would seem to be the most sensible option. 
This prescription is nonetheless
worthy of study, and is a ``continuous KK-mode
distribution approximation'' to any theory
that envelops the Standard Model into a noninteger extra
dimension $\epsilon$. 

We have in mind other applications of these ideas.
If the theories at lattice sites live in
$1+d$ continuous dimensions, then the full theory has
$1+d+\epsilon$ dimensionality. 
For example, with continuum $1+1$ fields
and very large $s$ we can construct a flat $4-\epsilon'$
dimensional effective theory. Such theories are classically
asymptotically free, but the log dependence on $s$ implies
that it would be difficult to construct a natural lattice
of the kind we have considered, since $s$ 
must be taken unnaturally large.  It is therefore of interest to
enlarge the space of recursively defined lattices to see
if natural $4-\epsilon$ dimensions make sense with very small $\epsilon$.

Yet another interesting possibility is to deconstruct the string
world-sheet.  Weyl invariance may be realized as
a discrete RG invariance on a latticized world-sheet 
with a continuum limit.
Then we may be able to find unusual generalizations of the string
theory to fractal worldsheets. The consequences for the Weyl
constraints on the target space may be interesting. 

The key result of this paper is that
the  renormalization
group is dual to geometry. The latter acts in space,
and the former acts in
theory space.  The RG can then be used to
reverse engineer unusual new geometries.
These fractal geometries exploit quantum mechanics
in their construction in a fundamental way. In fact,
these realize some of the recent
speculations about deconstruction as a means of reconstructing
spacetime \cite{georgi}.   Though we are far
from a  complete theory, \eg, one including 
gravity, we believe this is fertile
territory with potentially nontrivial implications beyond those
considered presently. 

\newpage 
\section{Appendix: Properties of the Vacuum Energy}

The vacuum energy integral is most readily computed
by differentiating {wrt} $\mu^2$. 
Define $\Gamma = c(\epsilon)(Z^{-2}N/\Lambda^\epsilon)V(\mu^2)$
where $V(\mu^2)$ is the Euclidean integral:
\beq
V(\mu^2) = \int \frac{d^{4+\epsilon}p}{(2\pi)^{4+\epsilon}}
\ln(p^2 + \mu^2)
\eeq
Then:
\beq
\frac{\partial}{\partial\mu^2}V(\mu^2) = 
\int \frac{d^{4+\epsilon}p}{(2\pi)^{4+\epsilon}}
\frac{1}{p^2 + \mu^2} = \frac{1}{(4\pi)^{2+\epsilon/2}}(\mu)^{2+\epsilon}
{\Gamma(-1-\epsilon/2)}
\eeq
Therefore, integrating wrt $\mu^2$,
noting the integral
vanishes for $\mu=0$ for the $4+\epsilon$ range
of interest:
\beq
V(\mu^2) = \frac{1}{(4\pi)^{2+\epsilon/2}}(\mu)^{4+\epsilon}
\frac{\Gamma(2-\epsilon/2)}{(2+\epsilon/2)(1+\epsilon/2)(\epsilon/2)(1-\epsilon/2)}
\eeq
The integral can be performed directly, without
differentiating, but is
more tedious.
In performing the integral this way we have made
two assumptions (1) The RG invariance holds for
the system under the integral sign; 
(2) the integral is Lorentz
invariant in $4+\epsilon$ dimensions. 
When these symmetries
are imposed ``under the integral'' we have
the RG invariant theory.  Such strong assumptions,
while the preferred interpretation of the theory, are
not necessary, however. 

It is useful to consider the integral before
we take the continuum limit. 
Suppose we do not take the $k\rightarrow \infty$
limit first and perform the integral with a cut-off.
Consider, from
eq.(3.33) and eq.(3.34) the integral
with cutoff $K=\sqrt{c}\Lambda$:
\beq
V_K'(\mu^2) = \int_0^K \frac{d^{4+\epsilon}p}{(2\pi)^{4+\epsilon}}
\frac{1}{(p^2 + \mu^2)}
\eeq
Here we have assumed that the finite
cut-off is Lorentz invariant
in $4+\epsilon$ dimensions.  We exponentiate
the denominator:
\beq
V_K'(\mu^2) = \int_0^\infty d\alpha 
\int_0^K \frac{d^{4+\epsilon}p}{(2\pi)^{4+\epsilon}} 
e^{-\alpha(p^2 + \mu^2)}
\eeq
The $p^2$ integral is then carried out, using
the solid angle in $d$ dimensions:
\be
\Omega_d = \frac{2\pi^{d/2}}{\Gamma(d/2)}
\ee
we have:
\beq
V_K'(\mu^2) =  \frac{1}{(2\pi)^{2+\epsilon/2}} \int_0^\infty d\alpha 
\left(\int_0^{\infty}- \int_{K^2}^{\infty}\right)  dp^2 (p^2)^{1+\epsilon/2} 
e^{-\alpha(p^2 + \mu^2)}
\eeq
The second term can be approximated:
\bea
V_K'(\mu^2) & = & V'(\mu^2)-
\frac{1}{(2\pi)^{2+\epsilon/2}} \int_0^\infty d\alpha 
(K^2)^{2+\epsilon/2} 
e^{-\alpha(K^2+\mu^2)} \nonumber \\
& \approx & V'(\mu^2)-
\frac{1}{(2\pi)^{2+\epsilon/2}} 
\frac{(K^2)^{2+\epsilon/2} }{K^2 + \mu^2}
\eea
hence, 
\bea
V_K(\mu^2) 
& \approx & V(\mu^2)-
\frac{1}{(2\pi)^{2+\epsilon/2}} 
\mu^2{(K^2)^{1+\epsilon/2} } +{\cal{O}}(K^2)^{2+\epsilon/2}
\eea
The  procedure of considering 
$V_K'(\mu^2)$ omitted the last quartically
divergent piece (when $\epsilon=0$), but keeps the
quadratically divergent piece. 

These additional terms are associated with
the UV cut-off on the theory, $\Lambda=K/\sqrt{c}$. Hence they are not RG invariant. 
These terms do not obstruct our measurement of the dimensionality
of the system. If the system has finite $k$, then the dimensionality
we compute from the finite
terms is ``effective,'' applying only near the threshold
of KK modes, $\sim M$.  The far UV spectrum can thus be
considerably modified by the finite $\Lambda$ effects which do
not respect the RG invariance.  

Clearly we would like to argue that the $4+\epsilon$
dimensional theory exists, and the $k\rightarrow\infty$
limit can be taken. In this case, such
terms can 
be viewed as the finite--$k$ regulator effect spoiling the symmetry of
RG invariance in the true theory, much the way a momentum-space cut-off spoils
gauge invariance. These terms vary under the RG transformation, so
they would be subtracted to define the symmetric theory, much
as gauge invariance forces the subtraction of the momentum-space
cut-off mass term in the vacuum polarization loop. It is simpler
to take the limit under the integral sign, since
the resulting integral is finite!

Casimir effects arise in principle
in the difference between
the continuous approximation
to the theory and the discretized sum. The form of these effects
depends upon how we define the theory, as described
above, and will generally have divergences associated
with finite UV cut-off's that are not RG
invariant.  However, when we define the theory
to be $4+\epsilon$
Lorentz invariant and take the UV cut-off to infinity
then the Casimir effects are vanishing.

Consider the original discrete vacuum energy expression:
\be
\Gamma = Z^{-2}\int \frac{d^4p}{(2\pi)^4} 
\sum_{n} \ln(p^2 + \omega_n^2 +\mu^2)
\ee
The Casimir effect is the difference between the continuum
integral approximation and the ``staircase''  
in the discrete sum.  This difference
is minimized by the best fit continuous
approximation to the staircase, 
but a leading residual contribution remains
involving the second derivative of the summand.
In our case the leading
Casimir correction to the continuous approximation
is:
\be
\delta\Gamma = \frac{1}{4} \frac{Z^{-2}c\Lambda^2}{N^2_k}
\int \frac{d^4p}{(2\pi)^4} \sum_{n} 
\left[\frac{2}{(p^2 + \omega_n^2+\mu^2)}
-  \frac{4\omega^2_n}{(p^2 + \omega_n^2+\mu^2)^2}\right]
\ee
Note the factor of ${c}\Lambda^2/N^2_k$  arising from 
$(\partial\omega_n/\partial n)^2$.

We rescale, and take the continuum Lorentz
invariant limit the integral-sum goes over to
a $4+\epsilon$ dimesnional integral. We have, by 
the Euclidean invariance,
$\omega^2_n = [\epsilon/(4+\epsilon)]p^2$.
Hence:
\be
\delta\Gamma = \half c(\epsilon)\frac{Z^{-2}N_k}{\Lambda^{\epsilon}}
\left(\frac{c\Lambda^2}{N^2_k}\right)
\int \frac{d^{4+\epsilon}p}{(2\pi)^{4+\epsilon}} \left[ 
\left(\frac{4-\epsilon}{4+\epsilon}\right)
\frac{p^2}{(p^2 +\mu^2)^2}
+
\frac{2\mu^2}{(p^2 +\mu^2)^2} \right]
\ee
The integral is readily performed:
\be
\delta\Gamma = \half (\epsilon)
\left(\frac{Z^{-2}N_k(\mu)^{4+\epsilon}}{(4\pi)^{2+\epsilon/2}
\Lambda^\epsilon}\right)
\left(\frac{c\Lambda^2}{N^2_k\mu^2}\right)
  {\Gamma(-1-\epsilon/2)}
 \left[ 
\left(\frac{4-\epsilon}{4+\epsilon}\right)
+ (1+\epsilon/2)\left(\frac{4+3\epsilon}{4+\epsilon}\right)
\right]
\ee
In this result we have the usual RG invariant
prefactor ${Z^{-2}N_k(\mu)^{4+\epsilon}}/{\Lambda^\epsilon} $.
We also have an additional factor of ${\Lambda^2}/{N^2_k\mu^2}$.
which can be written in terms of invariants $M$ and ${\mu}_r$ as:
${M^2}/{{N_k}^{2-2/\epsilon}{\mu}_r^2}$. As we take the UV limit
the number of sites grows as $N_k\rightarrow (s+1)s^{k}$, hence
the prefactor term vanishes in this limit. 
Higher order Casimir corrections will involve higher
powers of $N_k$ in the denominator of the prefactor.
The Casimir corrections
to the vacuum energy are thus vanishing  to all orders
when the theory is defined as a $4+\epsilon$ Lorentz
invariant continuum theory.

\vspace*{1.0cm}  
\noindent  
{\bf \Large \bf Acknowledgements}  
  
I wish to thank K. Dienes, A. Kronfeld and
Y. Nomura for useful discussions. 
Research was supported by the U.S.~Department of Energy  
Grant DE-AC02-76CHO3000.  
\frenchspacing  
  
\end{document}